\documentclass[11pt]{article}

\usepackage{amsmath,amsfonts}

\usepackage{amssymb}

\usepackage{lscape}

\usepackage[dvips]{epsfig}

\usepackage[dvips]{graphicx}

\begin{document}

\begin{center}

{\Large Enhanced diffraction by
a rectangular grating made of a negative phase--velocity (or negative index) material}

\bigskip

\emph{Ricardo A. Depine}\footnote{Corresponding Author. E-mail: rdep@df.uba.ar}\\
{Departamento de F\'{\i}sica, Facultad de Ciencias Exactas y Naturales,\\
Universidad de Buenos Aires, 1428 Buenos Aires, Argentina}\\
\medskip

\emph{Akhlesh Lakhtakia}\footnote{E-mail: akhlesh@psu.edu}\\
{Department of Engineering Science and Mechanics, Pennsylvania State
University, University Park, PA 16802--6812, USA}\\
\medskip

\emph{David R. Smith}\footnote{E-mail: drsmith@ee.duke.edu}\\
{Department of Electrical and Computer Engineering, Duke University, \\ Durham, NC 27708, USA}\\
\date{\today}

\end{center}

\bigskip

\noindent {\bf Abstract.}
The diffraction of electromagnetic plane waves by a rectangular grating formed
by discrete steps in the interface of a homogeneous, isotropic, linear, negative phase--velocity
(negative index) material with free space is studied using the semi--analytic C method. When a nonspecular diffracted order
is of the propagating type, coupling to that order
is significantly larger for a negative index material than for conventional material. The computed coupling strengths reported here 
 are in agreement with recent experiments, and
illustrate the role of evanescent fields localized at the grating interface in
producing this enhanced coupling.

\noindent PACS: {42.25.Fx, 78.20.Ci}

\noindent Keywords: diffraction, grating, negative phase velocity, negative refractive
index 

%------------------------------------------------------

% It is always \today, today,
%  but any date may be explicitly specified

\newpage

The study of electromagnetic fields received a major boost in 2001 with the
experimental confirmation of negative refraction by dielectric--magnetic
materials with negative real permittivity and negative real permeability 
\cite{smith}. When considered as linear, isotropic and homogeneous, these
materials possess a complex--valued refractive index whose real part is
negative, and are therefore often called \emph{negative index\/} (NI)
materials. Alternatively, these materials can be referred to as \emph{%
negative phase--velocity} (NPV) materials, as the phase velocity and the
time--averaged Poynting vectors therein are antiparallel. Other names are
also in circulation; but,  regardless of the name used, the common
observable phenomenon is negative refraction \cite{GE02,PGLKT03,HBC03}.

All experimental realizations of NPV materials thus far are as periodically
patterned composite materials. The unit cell comprises various arrangements
of conducting filaments in order to realize both dielectric and magnetic
response properties in the same frequency range. For such a material to be
considered as effectively homogeneous, the unit cell must be electrically
small~---~i.e., it must be considerably smaller than the free--space
wavelength as well as the wavelength in the material \cite{deH}. In the
wedge sample used by Shelby \emph{et al.\/} \cite{smith}, the unit cell size
was about one--sixth of the free--space wavelength. The finite size of the
unit cell meant that one of the two exposed surfaces of the sample was not
planar, but rather formed a rectangular grating \cite{May}. Because of this
grating, the specular (or zeroth--order) reflected/refracted plane wave 
\emph{could\/} be accompanied by nonspecular (higher--order) diffracted plane waves \cite%
{May}. 

Recently, using Ansoft's HFSS package, an electromagnetics equation solver
based on the finite--element method, Smith \emph{et al.\/} \cite{SRMVF}
examined the plane--wave response of a wedge having the same properties as
that used by Shelby \emph{et al.\/} \cite{smith}. One of the two exposed
surfaces was set up as a shallow rectangular grating. The simulation
revealed not only a zeroth--order transmitted plane wave but also,
unexpectedly, a strong first--order transmitted plane wave. The latter was
much stronger for a NPV wedge than for a wedge with identical dimensions but
made of a positive phase--velocity (PPV) material. Although higher--order
refracted plane waves were not experimentally observed by Shelby \emph{et
al.\/} \cite{smith}, the first--order was definitively present in later
experiments \cite{HBC03,SRMVF}.

While full--wave simulations confirm the phenomenon of enhanced diffraction
at the NPV grating, the numerical results are not of use in determining the
physical origin of the effect. \ An alternative approach is furnished by mathematical
treatments of diffraction by gratings based on analytic field expansions. These treatments have
continued to develop over the last hundred years, and have now acquired a
considerable degree of sophistication. We apply one of these methods here,
the so--called C method, both to independently verify the simulation results
obtained by the finite--element method and to provide further insight as to
the underlying mechanism \cite{SRMVF}. The C method, originally developed
for dielectric gratings \cite{LiChande}, was modified to handle
dielectric--magnetic gratings \cite{DLprsa}.

\begin{figure}[!ht]
%\vskip -0.5cm
\begin{center}
\begin{tabular}{c}
\includegraphics[width=5cm]{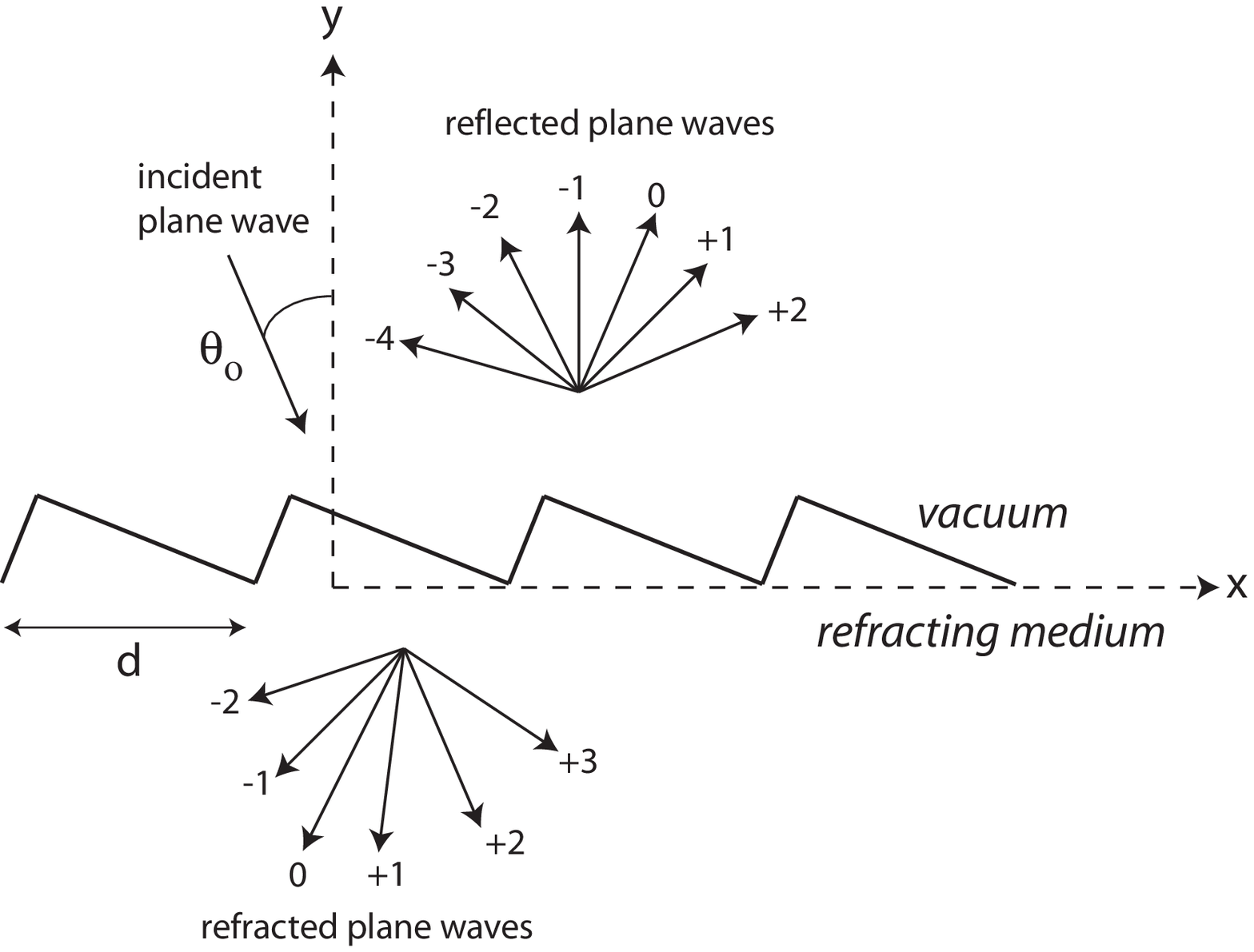}
\end{tabular}
\end{center}
%\vskip -1.5cm
\caption[example]{Schematic of the diffraction problem solved. The refracted plane waves are shown as if the refracting medium is of
the NPV type. The specular reflected and refracted orders are  denoted by $n=0$, while nonspecular orders are denoted by $n\ne 0$.%
The inset shows the shape of the grating profile used by Smith {\em et al.\/} \cite{SRMVF} as well as for Figures \ref{fig2}--\ref{fig4}.
\label{fig1}} \end{figure}

In a rectangular coordinate system $(x,y,z)$, we consider the periodically
corrugated boundary $y=g(x)=g(x+d)$ between vacuum and a homogeneous,
isotropic, dielectric--magnetic material, with $d$ being the corrugation
period, as shown in Figure \ref{fig1}. The region $y>g(x)$ is vacuous,
whereas the medium occupying the region $y<g(x)$ is characterized by
complex--valued scalars $\epsilon _{2}=\epsilon _{2R}+i\epsilon _{2I}$ and $%
\mu _{2}=\mu _{2R}+i\mu _{2I}$, such that $\epsilon _{2I}\geq 0$ and $\mu
_{2I}\geq 0$. The refracting medium is assumed to be \emph{effectively\/}
homogeneous at the angular frequency of interest. A linearly polarized
electromagnetic plane wave is incident on this boundary from the region $%
y>g(x)$ at an angle $\theta _{0}$, $(|\theta _{0}|<\pi /2)$, with respect to
the $y$ axis.

Let the function $f(x,y)$ represent the $z$--directed component of the total
electric field for the $s$--polarization case, and the $z$--directed
component of the total magnetic field for the $p$--polarization case \cite%
{BW80}. Outside the corrugations, $f(x,y)$ is rigorously represented by
Rayleigh expansions \cite{donR} as 
\begin{eqnarray}
f(x,y) &=&\exp \left[ i\,(\alpha _{0}x-\beta _{0}^{(1)}y)\right] +  \nonumber
\\
&&\sum_{n=-\infty }^{+\infty }\rho _{n}\,\exp \left[ i\,(\alpha _{n}x+\beta
_{n}^{(1)}y)\right] \,,  \nonumber \\
&&\qquad \qquad y>\mbox{max}\,g(x)\,\,,  \label{f1}
\end{eqnarray}%
and 
\begin{eqnarray}
f(x,y) &=&\sum_{n=-\infty }^{+\infty }\tau _{n}\,\exp \left[ i\,(\alpha
_{n}x-\beta _{n}^{(2)}y)\right] \,,  \nonumber \\
&&\qquad y<\mbox{min}\,g(x)\,\,.  \label{f2}
\end{eqnarray}%
Here, $\left\{ \rho _{n}\right\} _{n=-\infty }^{\;\;\;\;+\infty }$ and $%
\left\{ \tau _{n}\right\} _{n=-\infty }^{\;\;\;\;+\infty }$ are scalar
coefficients to be determined; and 
\begin{equation}
\left. 
\begin{array}{ll}
\alpha _{0}=\frac{\omega }{c}\,\sin \theta _{0}\,,\,\,\alpha _{n}=\alpha
_{0}+{2n\pi }/{d} &  \\[5pt] 
\beta _{n}^{(1)}=\sqrt{\frac{\omega ^{2}}{c^{2}}-\alpha _{n}^{2}}%
\,,\,\,\beta _{n}^{(2)}=\sqrt{\frac{\omega ^{2}}{c^{2}}\epsilon _{2}\,\mu
_{2}-\alpha _{n}^{2}} & 
\end{array}%
\right\} \,,
\end{equation}%
where $c$ is the speed of light in vacuum and $\omega $ is the angular
frequency. Note that $\beta _{n}^{(1)}$ is either purely real or purely
imaginary; and the conditions $\mathrm{Re}\left[ \beta _{n}^{(1)}\right]
\geq 0$ and $\mathrm{Im}\left[ \beta _{n}^{(1)}\right] \geq 0\,\forall n$
are appropriate for plane waves in the vacuous half--space $y>\mbox{max}%
\,g(x)$. The refracted plane waves must attenuate as $y\rightarrow -\infty $%
, imposing the condition $\mathrm{Im}\left[ \beta _{n}^{(2)}\right] >0$.
Fulfillment of this condition automatically fixes the sign of $\mathrm{Re}%
\left[ \beta _{n}^{(2)}\right] $, regardless of the signs of $\epsilon _{2R}$
and $\mu _{2R}$.

After implementing the C method on a computer, the coefficients $%
\left\{\rho_n\right\}_{n=-\infty}^{\;\;\;\;+ \infty}$ and $%
\left\{\tau_n\right\}_{n=-\infty}^{\;\;\;\;+ \infty}$ are determined.
Diffraction efficiencies 
$
e_{n}^{\rho} = ({\mathrm{Re}\left[\beta^{(1)}_n\right]}/{\beta^{(1)}_0})%
\,\vert\rho_n\vert^2 $
%\label{ernr}
are defined for the reflected orders. If dissipation in the refracting
medium is small enough to be ignored, diffraction efficiencies 
$
e_{n}^\tau = ({\mathrm{Re}\left[\beta^{(2)}_n\right]}/{\sigma
\,\beta^{(1)}_0})\,\vert\tau_n\vert^2 $
% \label{ernt}
are defined for the refracted orders, where either $\sigma=\mu_2$ ($s$%
--polarization) or $\sigma =\epsilon_2$ ($p$--polarization).

The grating of Shelby \emph{et al.\/} \cite{smith} has a rectangular
profile, shown in Figure \ref{fig1}, with the long and the short sides in
the ratio $3:1$ and period $d=15.81$~mm. To simulate this profile for the C
method, $g(x)$ was replaced by the truncated Fourier sum $\sum_{n=0}^{10}
\gamma_n\sin(2\pi nx/d+\varphi_n)$, which was found adequate to represent
the grating profile.

\begin{figure}[!ht]
%\vskip -0.5cm
\par
\begin{center}
\begin{tabular}{c}
\includegraphics[width=3.5cm]{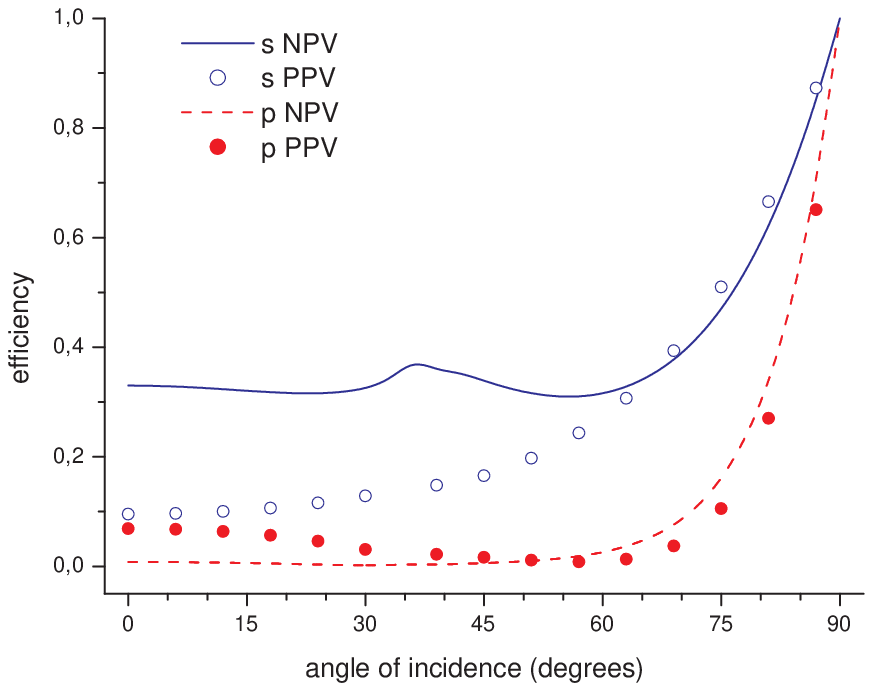} %
\includegraphics[width=3.5cm]{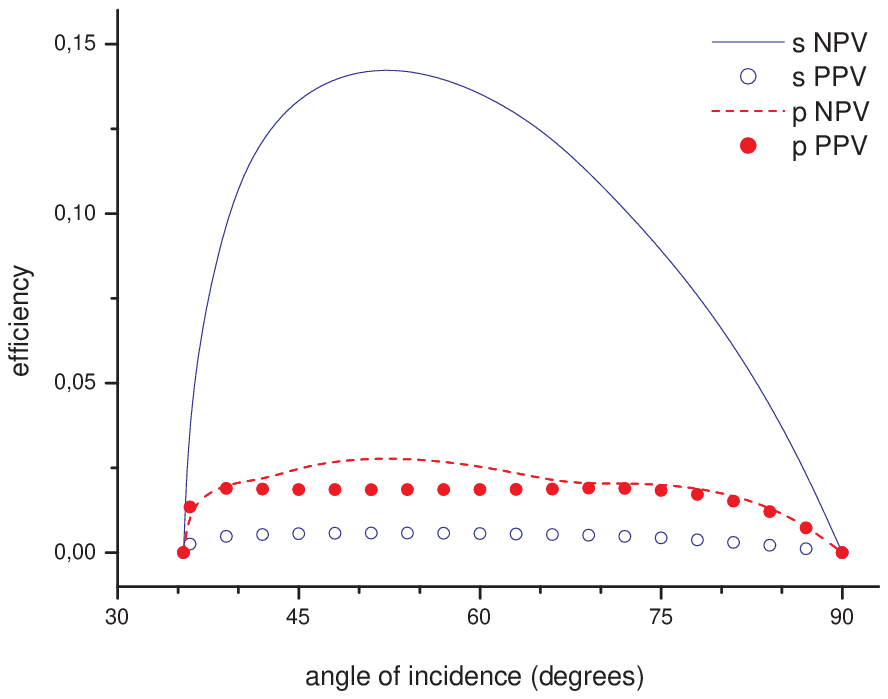} \\ 
\includegraphics[width=3.5cm]{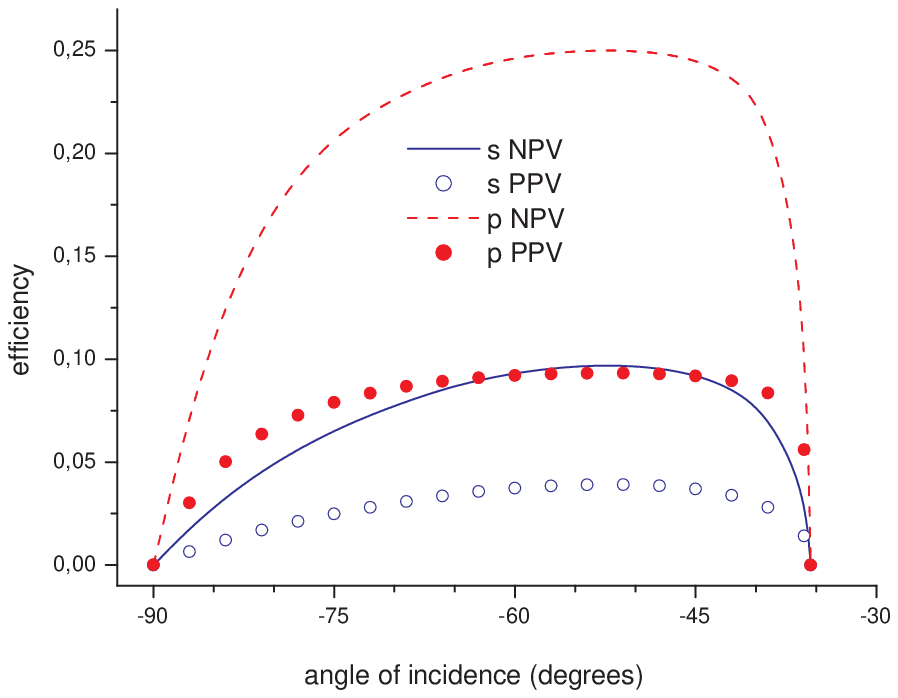}%
\end{tabular}%
\end{center}
\par
%\vskip -1.5cm
\caption[example]{Computed diffraction efficiencies as functions of the
angle of incidence $\protect\theta_0$, when $\protect\omega d/c=2\protect\pi%
/1.58$, for both $p$-- and $s$--polarized plane waves. The refracting medium
is either of the PPV ($\protect\epsilon_2=5+i0.01,\,\protect\mu_2=1+i0.01$)
or the NPV ($\protect\epsilon_2=-5+i0.01,\,\protect\mu_2=-1+i0.01$) type.
(a) $e_0^\protect\protect\rho$; (b) $e_{-1}^\protect\protect\rho$; (c) $%
e_{+1}^\protect\protect\rho$.%
}
\label{fig2}
\end{figure}

In Figure \ref{fig2}, we show the diffraction efficiencies obtained as
functions of $\theta _{0}$ at a frequency of 11.75 GHz, the highest
frequency used by Smith \emph{et al.\/} \cite{SRMVF}, which corresponds to $%
\omega d/c=2\pi /1.58$. The refracting medium is of either the the NPV ($%
\epsilon _{2}=-5+i0.01,\,\mu _{2}=-1+i0.01$) or the PPV ($\epsilon
_{2}=5+i0.01,\,\mu _{2}=1+i0.01$) type. Calculations were made for both the $%
s$-- and the $p$--polarization incidence conditions. As confirmed by
perturbation analysis \cite{NPV-oc}, in this case the transformation $%
\left\{ \epsilon \rightarrow -\epsilon ^{\ast },\,\mu \rightarrow -\mu
^{\ast }\right\} $ does not greatly affect $e_{0}^{\rho }$ (Fig. \ref{fig2}%
a), except at low $|\theta _{0}|$. In contrast, Figs. \ref{fig2}b and \ref%
{fig2}c show that the nonspecular reflection efficiencies $e_{\pm 1}^{\rho }$%
, are significantly affected by the type of the refracting medium. In
particular, the interplay between the polarization state and the angle of
incidence leads to (i) very little difference for the PPV grating between
the two polarizations; (ii) considerable difference for the NPV grating
between the two polarizations, and (iii) a shift of the difference between
PPV and NPV gratings from one polarization to the other as the sign of the
angle of incidence changes. As the refracting medium is dissipative,
diffraction efficiencies for the refracted orders cannot be defined.

\begin{figure}[!ht]
%\vskip -0.5cm
\par
\begin{center}
\begin{tabular}{c}
\includegraphics[width=3.5cm]{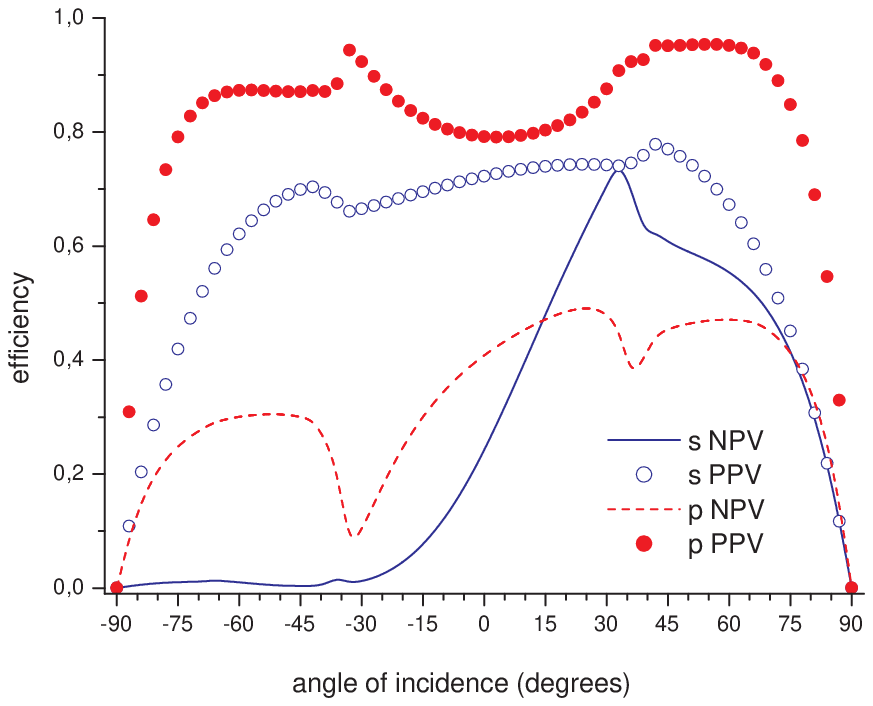} %
\includegraphics[width=3.5cm]{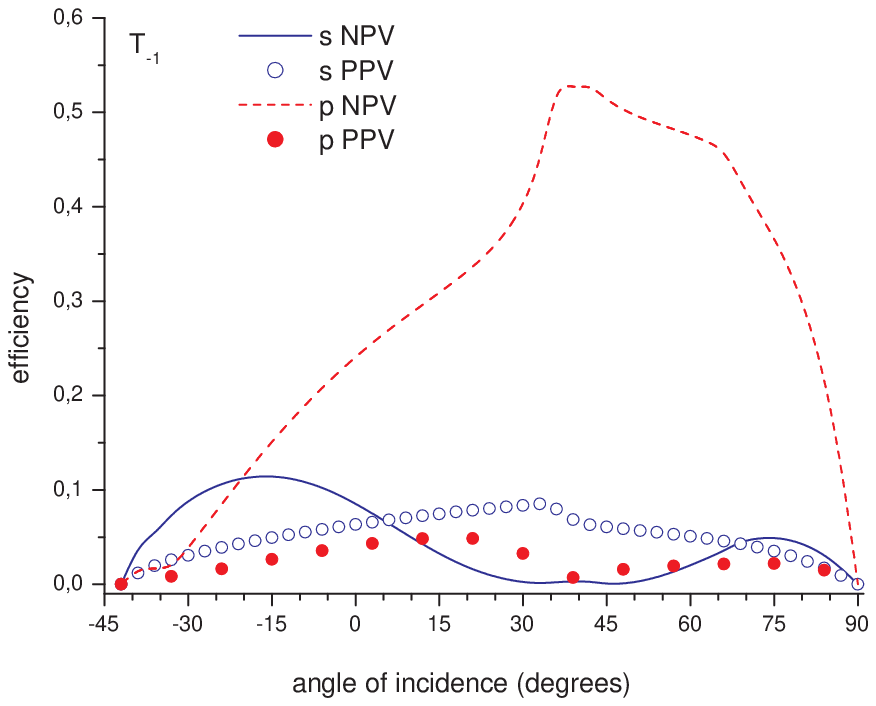} \\ 
\includegraphics[width=3.5cm]{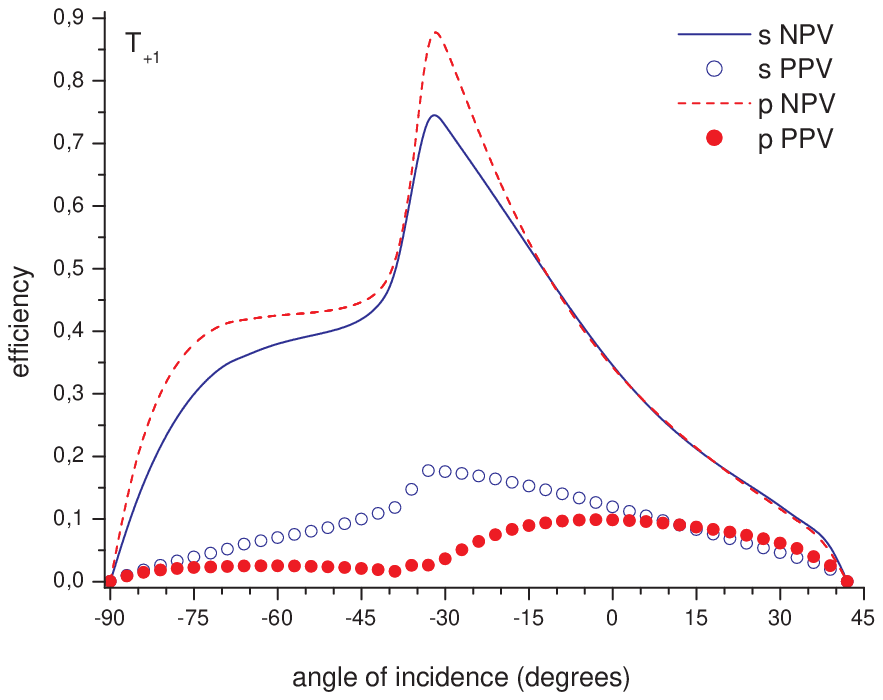}%
\end{tabular}%
\end{center}
\par
%\vskip -1.5cm
\caption[example]{Computed diffraction efficiencies as functions of the
angle of incidence $\protect\theta_0$, when $\protect\omega d/c=2\protect\pi%
/1.58$, for both $p$-- and $s$--polarized plane waves. The refracting medium
is either of the PPV ($\protect\epsilon_2=5,\,\protect\mu_2=1$) or the NPV ($%
\protect\epsilon_2=-5,\,\protect\mu_2=-1$) type. (a) $e_0^\protect\protect%
\tau$; (b) $e_{-1}^\protect\protect\tau$; (c) $e_{+1}^\protect\protect\tau$. 
}
\label{fig3}
\end{figure}

To study refraction efficiencies explicitly, we made the refracting material
nondissipative ($\epsilon_2=\pm 5,\,\mu_2=\pm 1$), while the other
parameters were kept the same. The higher--order reflection efficiency
curves display the same qualitatitive behaviors as in Figure \ref{fig2}, and
are therefore not shown in Figure \ref{fig3}. All refraction efficiencies
are greatly affected by the type of the refracting medium, as becomes clear
from Figs. \ref{fig3}a--c. In particular, the coupling of power into the $%
n=\pm 1$ refracted orders (Figs. \ref{fig3}b and c) for the NPV grating is
much larger than for a PPV grating, a fact that again is in agreement with
the results of Smith \emph{et al.} \cite{SRMVF}.

\begin{figure}[!ht]
%\vskip -0.5cm
\par
\begin{center}
\begin{tabular}{c}
\includegraphics[width=3.5cm]{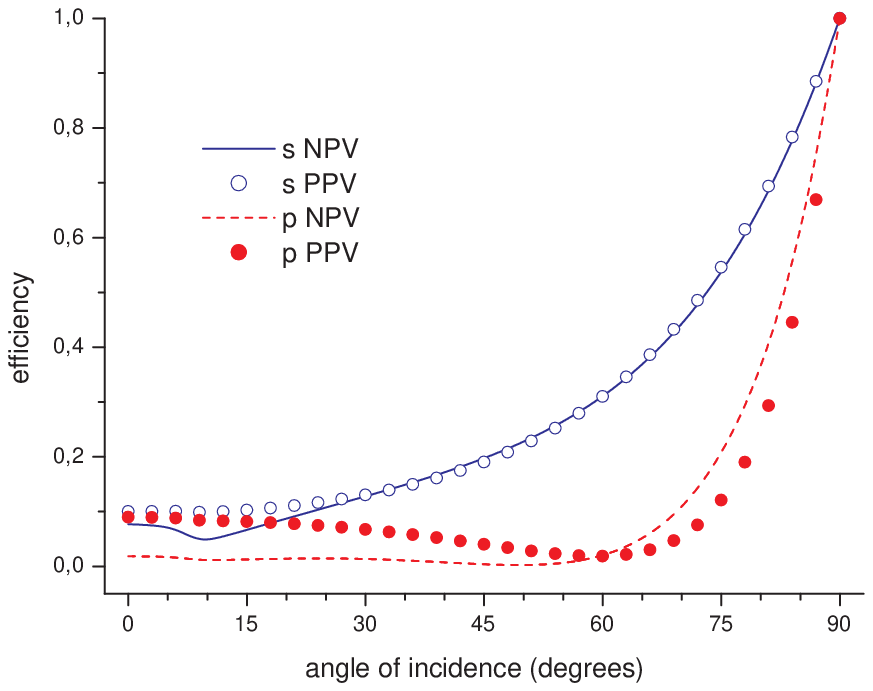} %
\includegraphics[width=3.5cm]{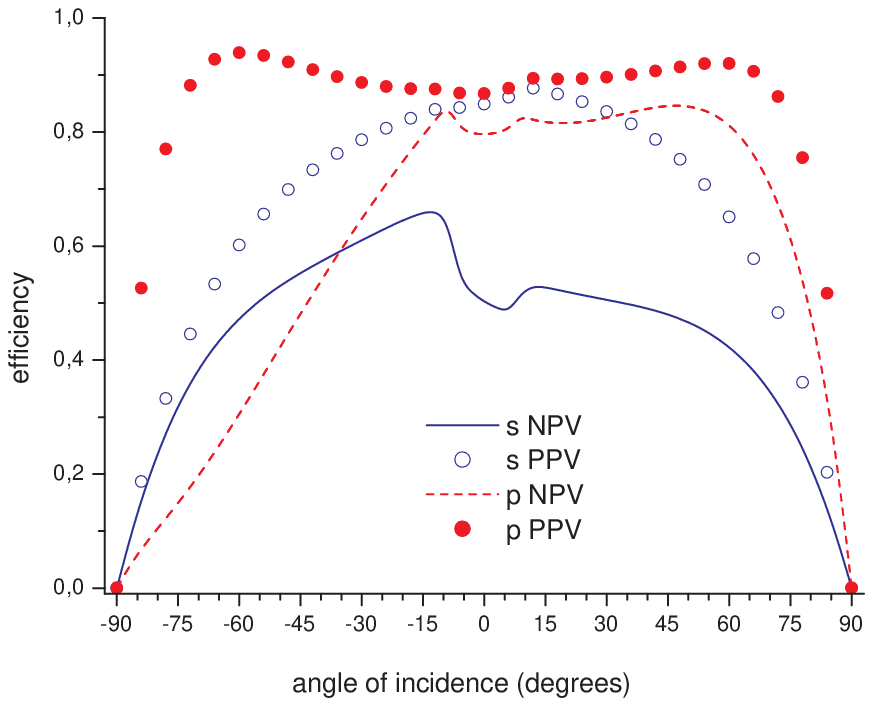} \\ 
\includegraphics[width=3.5cm]{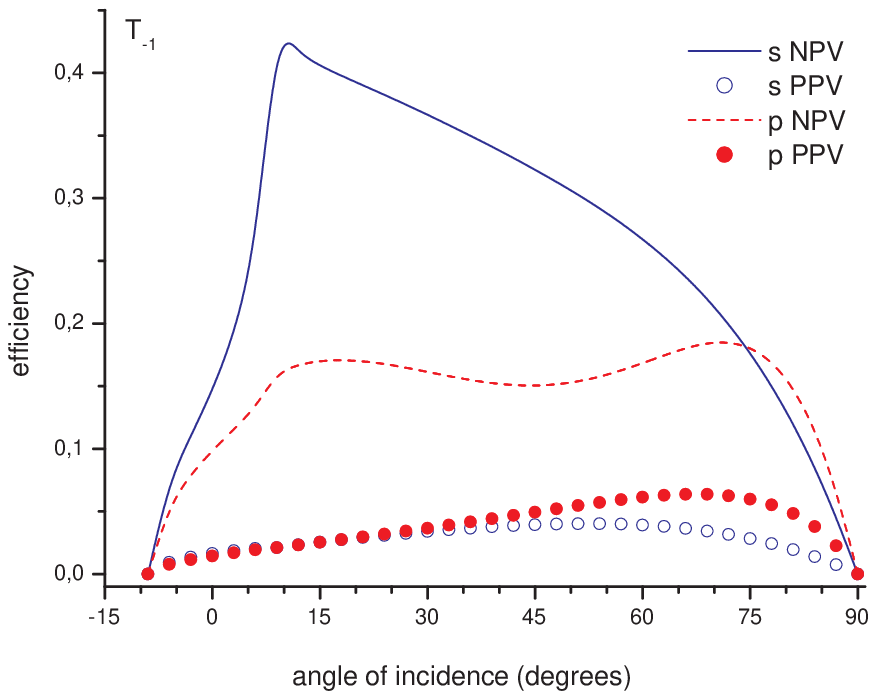} %
\includegraphics[width=3.5cm]{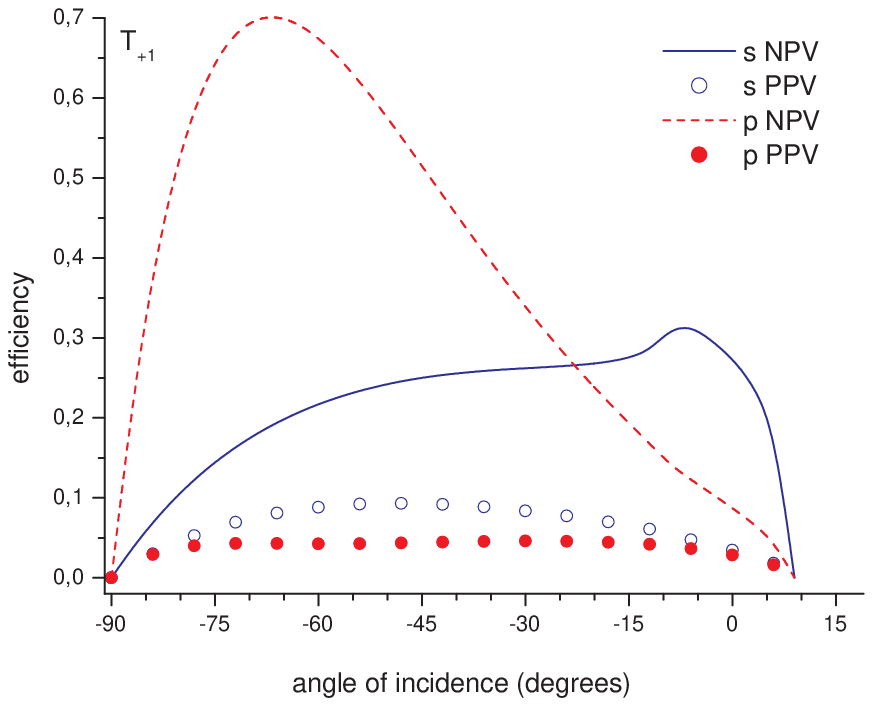}%
\end{tabular}%
\end{center}
\par
%\vskip -1.5cm
\caption[example]{Computed diffraction efficiencies as functions of the
angle of incidence $\protect\theta_0$, when $\protect\omega d/c=2\protect\pi%
/2.087$, for both $p$-- and $s$--polarized plane waves. The refracting
medium is either of the PPV ($\protect\epsilon_2=5,\,\protect\mu_2=1$) or
the NPV ($\protect\epsilon_2=-5,\,\protect\mu_2=-1$) type. (a) $e_0^\protect%
\protect\rho$; (b) $e_0^\protect\protect\tau$; (c) $e_{-1}^\protect\protect%
\tau$; (d) $e_{+1}^\protect\protect\tau$. }
\label{fig4}
\end{figure}

A similar conclusion is drawn from Figure \ref{fig4}, for which the
efficiency curves were computed at a frequency of 9.0 GHz (the smallest
frequency used by Smith \emph{et al.\/} \cite{SRMVF}), the other parameters
being the same as in Figure \ref{fig3}. Whereas the zeroth--order (Fig. \ref%
{fig4}a, the only propagating order reflected at this frequency) reflection
efficiency is much less affected by the type of the refracting medium than
in our previous examples, the refraction efficiencies depend strongly on the
refracting medium being of the PPV or the NPV type.

The increased coupling into higher--order  diffracted plane waves represents
an important distinction between the behaviors of PPV and NPV materials, and
confirms that surface periodicity plays a much more significant role in the
latter \cite{SRMVF}. The physical origin of the enhanced higher--order
reflection efficiencies for the NPV grating can be understood in an approximate manner
by examining the reflection efficiencies for shallow gratings. In general, a grating couples the
specular coefficients ($\rho_0$ and $\tau_0$) with all nonspecular
coefficients ($\rho_n$ and $\tau_n$, $\vert n\vert > 0$) \cite{May},
thereby coupling $e_0^\rho$ with all $e_n^\rho$, $\vert n\vert > 0$. All coefficients  for a grating of arbitrary profile are obtained together
by some numerical technique.

In the limit of the grating profile becoming planar, the reflection coefficients (and, therefore, the efficiencies)
coincide with those for a planar interface. The introduction of shallow corrugations only
slightly perturbs the coefficients \cite{Elson}, and we may then write
\begin{equation}
\label{appr}
\rho _{n}\simeq -\,\frac{\beta _{n}^{(2)}-\sigma \beta _{n}^{(1)}}{\beta _{n}^{(2)}+\sigma
\beta _{n}^{(1)}}\,.
\end{equation}
In the limit of small  dissipation,
$\beta
_{n}^{(2)}$ is real--valued for propagating orders in the refracting medium. As this medium transforms
from PPV to NPV, $\beta
_{n}^{(2)}$ changes sign along with $\sigma$, and  the magnitude of $\rho_n $ is thus
unchanged.
But for evanescent (nonspecular) orders, $\beta
_{n}^{(2)}$ is imaginary, and thus always positive in accordance with
causality. Hence, for nonspecular orders,  the reflection coefficients are inverted as the refracting
medium is transformed from PPV to NPV.   The evanescent orders  then,
which play a minor role for  the PPV grating, mediate a much larger
interaction between the specular and the nonspecular modes for the NPV grating.

Equation (\ref{appr}) shows that there is the possibility of a pole for one or more reflection orders. A pole indicates
the presence of a bound surface mode, identical to the surface plasmon that occurs at the interface with a metal
at optical wavelengths. At least in a perturbative sense, the enhanced diffraction by a NPV grating has the same origin as 
other plasmon--mediated effects, including enhanced light transmission \cite{Ebb} and the  ``perfect lens" effect \cite{PenSmi}.

The detailed treatment presented here of diffraction from a NPV grating
reveals enhanced diffraction, in agreement with recent
experiments. As NPV materials are considered for various applications, the
results found here indicate that great care is necessary in describing the surfaces of
NPV materials. In particular, the recent
artificially structured metamaterials are based on periodic cells, whose
inherent periodicity can lead to a quite significant surface nonhomogeneity, which may assume even greater
importance for surfaces that are not nominally planar.

\vskip 0.5cm \noindent \textbf{Acknowledgments.} R.A.D. acknowledges partial
support from CONICET (Grant: ANPCYT-BID 802/OC-AR03-04457), and A.L.   from
the Mercedes Foundation.

%-------------
\newpage
%-------------

%-------------
\newpage
%-------------

%-------------
\newpage
%-------------

%-------------
\newpage
%-------------

\end{document}